\documentclass[]{article}
\usepackage{emulateapj}
\usepackage{graphicx}

\advance \voffset by -0.50cm\relax
\def\msun{{\rm ~M}_{\odot}}
\def\rsun{{\rm ~R}_{\odot}}

\def\mpy{{\rm ~M}_{\odot} {\rm ~yr}^{-1}}

\begin{document}

\title{The most massive progenitors of neutron stars: CXO J164710.2-455216}

 \author{Krzysztof Belczynski\altaffilmark{1,2,3}, 
         Ronald E.\ Taam\altaffilmark{4,5}}

 \affil{
     $^{1}$ Los Alamos National Laboratory, 
            CCS2/ISR1 Group, P.O. Box 1663, MS D466,
            Los Alamos, NM 87545\\
     $^{2}$ Oppenheimer Fellow\\
     $^{3}$ New Mexico State University, Dept of Astronomy,
            1320 Frenger Mall, Las Cruces, NM 88003\\
     $^{4}$ Northwestern University, Dept of Physics \& Astronomy,
            2145 Sheridan Rd, Evanston, IL 60208\\
     $^{5}$ ASIAA/National Tsing Hua University - TIARA, Hsinchu, Taiwan\\
     kbelczyn@nmsu.edu, r-taam@northwestern.edu}

\begin{abstract}
The evolution leading to the formation of a neutron star in the very young Westerlund 1 
star cluster is investigated. The turnoff mass has been estimated to be $\sim 35 \msun$, 
indicating a cluster age $\sim 3-5$ Myr. The brightest X-ray source in the cluster, 
CXO J164710.2-455216, is a slowly spinning ($10$ s) single neutron star and potentially a 
magnetar. Since this source was argued to be a member of the cluster, the neutron star 
progenitor must have been very massive ($M_{\rm zams} \gtrsim 40 \msun$) as noted by Muno 
et al. (2006). Since such massive stars are generally believed to form black holes (rather 
than neutron stars), the existence of this object poses a challenge for understanding 
massive star evolution.  We point out while single star progenitors below $M_{\rm zams} 
\lesssim 20 \msun$ form neutron stars, binary evolution completely changes the progenitor
mass range.  In particular, we demonstrate that mass loss in Roche lobe overflow enables 
stars as massive as $50-80 \msun$, under favorable conditions, to form neutron stars. If 
the very high observed binary fraction of massive stars in Westerlund 1 ($\gtrsim 70\%$) is 
considered, it is natural that CXO J164710.2-455216 was formed in a binary which was 
disrupted in a supernova explosion such that it is now found as a single neutron star. Hence, 
the existence of a neutron star in a given stellar population does not necessarily place 
stringent constraints on progenitor mass when binary interactions are considered. It is 
concluded that the existence of a neutron star in Westerlund 1 cluster is fully consistent 
with the generally accepted framework of stellar evolution.
\end{abstract}


\section{Introduction}

The recent detection of the variable X-ray source, CXO J164710.2-455216, located 1.'6 from
the core of the young open cluster, Westerlund 1, by Muno et al (2006) can be considered key to
understanding the
formation history and nature of compact objects in young massive stellar clusters. CXO
J164710.2-455216 is the brightest X-ray source in the cluster, radiating at a luminosity of
$\sim 3 \times 10^{33} (D/5 kpc)^2$ ergs s$^{-1}$ in the 0.5-8 keV energy band and
characterized by a pulse period of 10.6 s.  Based on subsequent XMM-Newton observations,
Israel et al. (2007) and Muno et al. (2007) showed that the source increased its luminosity
to $\sim 10^{35} (D/5 kpc)^2$ ergs s$^{-1}$ and exhibited a rapid spin down after an outburst,
exhibiting properties that are very reminiscent of highly magnetized neutron stars known as
magnetars.

Previous spectroscopic and photometric observations of Westerlund 1 were carried out by Clark
et al. (2005), who showed that the population of massive stars could be understood if the cluster
age was $4 \pm 1$ Myr (see also Crowther et al. 2006), a result which is also consistent with
a more recent age determination
based on an analysis of its intermediate and low mass stellar content (Brandner et al. 2008).
For an assumed solar metallicity, a lower limit for the progenitor mass of CXO J164710.2-455216
was estimated to be $\sim 35 \msun$. The reconstruction of the evolutionary history of CXO
J164710.2-455216 is highly desirable in placing limits on its progenitor, however, it is
dependent upon whether it evolved in isolation as a single star or as a member of an interacting
binary system.  We note that the recent studies by Clark et al. (2008) indicate a high binary
fraction, $\gtrsim 70\%$, for the massive Wolf Rayet star population in Westerlund 1 and that a
binary evolutionary channel may be very likely.

It is generally inferred from core collapse simulations that single stars more massive 
than $\sim 20-25 \msun$ form black holes (e.g., Fryer 1999). Stars 
more massive than $\sim 20-25 \msun$ in binaries are also expected to form black holes 
in order to explain the Galactic black hole transient systems and to satisfy nucleosynthetic 
constraints (e.g., Maeder 1992; Portegies Zwart, Verbunt \& Ergma 1997; Kobulnicky \& Skillman 
1997; Ergma \& van den Heuvel 1998).  However, a number of studies have been pointed out 
that binary interactions 
can significantly increase the neutron star/black hole mass formation limit. For example, 
Van den Heuvel \& Habets (1984) studied the binary LMC X-3 containing a neutron star with a 
massive companion, and found that stars as massive as $40 \msun$ may form neutron stars. 
Kaper et al. (1995) further increased this limit, estimating a mass of $50 \msun$ based on the 
analysis of a binary Wray 977, although it was later demonstrated that such a system could be 
explained with the much lower mass limit of $26 \msun$ (Wellstein \& Langer 1999). However, 
Wellstein \& Langer (1999) reemphasized that, in general, the critical initial mass limits for 
neutron star/black hole formation in binary systems may be quite different than for single
stars and can be as high as $100 \msun$.  We point out that, in this regard, Brown et al. 
(2001) have noted that the final mass of the compact object may depend upon whether the evolved 
core was exposed before or after central helium burning.  More recently, Fryer et al. (2002) 
demonstrated that a star of $60 \msun$ that was stripped of its envelope in a binary system 
can end up as a neutron star or a massive black hole, depending on the adopted wind mass loss 
rates during the Wolf-Rayet stage.  These studies, taken as a whole, indicate that binary 
interactions can lead to a range of initial stellar masses separating the formation of neutron stars 
from black holes rather than a single limiting mass.  We use this concept to examine the 
existence of a neutron star in the very young environment of Westerlund 1. 

In this paper, we report on the formation history of neutron stars from very massive progenitors
($\gtrsim 35 \msun$) both as a single star and as a component of an interacting binary system
to determine the conditions under which a neutron star in a young cluster such as Westerlund 1
can form. In the next section, we outline the basic assumptions and input parameters of our
numerical model as based on the {\tt StarTrack} binary population synthesis code (Belczynski et al.
2008). The numerical results are presented for the binary star and single star evolutionary channels
in \S 3.  The implications of these results for stellar and binary evolution of
massive stars, as applied to Westerlund 1, are discussed in the last section.

\section{Population Synthesis Model}

Our population synthesis code, {\tt StarTrack},  was initially developed to 
study double compact object mergers in the context of $\gamma$ ray burst progenitors
(Belczynski, Bulik \& Rudak 2002b) and gravitational-wave inspiral sources 
(Belczynski, Kalogera, \& Bulik 2002a). In recent years {\tt 
StarTrack} has undergone major updates and revisions in the physical treatment 
of various binary evolution phases, and especially the mass transfer phases. 
The new version has already been tested and calibrated against observations and 
detailed binary mass transfer calculations (Belczynski et al.\ 2008).  It  has 
been used in various applications (e.g., Belczynski \& Taam 2004; Belczynski et 
al.\ 2004; Belczynski, Bulik \& Ruiter 2005; Belczynski et al. 2006; Belczynski 
et al.\ 2007). The physics updates that are most important for compact object 
formation and evolution include: a full numerical approach for the orbital evolution 
due to tidal interactions, calibrated using high mass X-ray binaries and open 
cluster observations, a detailed treatment of mass transfer episodes fully 
calibrated against detailed calculations with a stellar evolution code, updated 
stellar winds for massive stars, and the latest determination of the natal
kick velocity distribution for neutron stars (Hobbs et al.\ 2005).  For helium 
star evolution, which is of a crucial importance for the formation of double 
neutron star binaries (e.g., Ivanova et al.\ 2003; Dewi \& Pols 2003), we 
have applied a treatment matching closely the results of detailed evolutionary 
calculations.  If the helium star fills its Roche lobe, the systems are examined 
for the potential development of a dynamical instability, in which case they 
are evolved through a common envelope (CE) phase, otherwise a highly non-conservative mass
transfer ensues. We treat CE events using the energy formalism (Webbink 1984),
where the binding energy of the envelope is determined from a set of He star 
models calculated with the detailed evolutionary code by Ivanova et al.\ (2003). 
In case the CE is initiated by a star crossing the Hertzsprung gap (HG) we 
assume a merger and abort further binary evolution. This is due to the fact 
that there is no clear core-envelope boundary (and no entropy jump as for 
more evolved stars) in the interior structure of HG donors to facilitate the
formation of a remnant binary system. As a consequence, a large decrease in 
the formation efficiency of close double compact binaries results (Belczynski 
et al. 2007). Rejuvenation is considered in detail following the method presented 
by Tout et al. (1997). For a detailed description of the revised code we refer 
the reader to Belczynski et al.\ (2008).

\section{Results}

\subsection{Binary star progenitor}

We have calculated the evolution of $2 \times 10^6$ massive binaries with our 
standard evolutionary model (Belczynski et al. 2008) for solar metallicity 
($Z=0.02$) and an initial mass function (IMF) with an exponent of $-2.7$ for primaries 
in the mass range $6 < M_{\rm zams,1} < 150 \msun$. The secondaries ($4 < M_{\rm zams,2} < 150 
\msun$)  were chosen from a flat mass ratio ($q=M_{\rm zams,2}/M_{\rm zams,1}$) 
distribution. The initial separations were chosen from the distribution flat in
logarithm (i.e., $\propto 1/a$) limited by the maximum separation of $10^5
\rsun$, while eccentricities were chosen from a thermal-equilibrium distribution 
($2e$). 

As a result we have obtained populations of compact objects, i.e., white
dwarfs, neutron stars and black holes, both single and in binaries. Single
compact objects were formed upon disruption of a given binary during a supernova
explosion either through a natal kick, mass loss or both. We have used the 
kick distribution presented by Hobbs et al. (2005) represented by a single
Maxwellian with $\sigma=265$ km s$^{-1}$. We employed the above distribution 
in its original form for neutron stars, while the kick velocity was decreased 
for black holes due to the fall back (e.g., Fryer \& Kalogera 2001).   

The stellar models employed in our population synthesis (Hurley, Pols \&
Tout 2000) result in an age of Westerlund 1 of 5 Myr: this is a main
sequence lifetime of a $35 \msun$ star. Actual fits of single stellar models 
to over 50 massive stars in Westerlund 1 resulted in an age estimate of 
$4 \pm 1$ Myr (e.g., Clark et al. 2005; Brandner et al. 2007). We adopt an age of 
5 Myr, although we note that any estimate based on single star models is highly 
uncertain in the case of a cluster like Westerlund 1 with the very high content 
of binaries ($\gtrsim 70\%$; Clark et al. 2008). At 5 Myr our calculations reveal 
a population of single neutron stars that originate from disrupted binaries. 

{\em Evolution. }
In the following we briefly describe a typical evolution leading to the formation 
of a very young solitary pulsar that originates from a disrupted binary. The evolution 
starts on the ZAMS with two massive stars: $M_{\rm zams,1}=65 \msun$ 
and $M_{\rm zams,2}=44 \msun$, on a rather close $a=220 \rsun$ and eccentric 
orbit: $e=0.65$. The primary initiates Roche lobe overflow (RLOF) at $t=3.3$ Myr while still 
on the main sequence (main sequence lifetime of an unaffected single star of that mass is 
$\tau_{\rm ms}=3.9$ Myr, and it forms a He core of $M_{\rm He}=18 \msun$). 
At the onset of mass transfer (Case A) the orbit is circularized ($a=80 \rsun$, $e=0$) and 
the component masses (depleted by stellar winds) are $M_{\rm 1}=52 \msun$ and 
$M_{\rm 2}=41 \msun$.  The mass transfer is stable and proceeds on the nuclear 
timescale of the primary with rates ranging from  $2 - 8 \times 10^{-5} \mpy$. 
We assume non-conservative evolution in which half of the mass lost by primary 
is accreted onto secondary, while the other half leaves the system, carrying away 
orbital angular momentum. At the end of the primary main sequence phase we note the mass 
ratio has reversed ($M_{\rm 1}=31 \msun$ and $M_{\rm 2}=49 \msun$) and the orbit has 
expanded ($a=100 \rsun$).  The significant mass loss (RLOF) on the main sequence leads to 
the decrease of the central temperature, reducing the rate of hydrogen (H) burning in the core 
of the primary. As a result, the primary forms a smaller core at the end of the main 
sequence phase (new $M_{\rm He}=10 \msun$) and, thus, will form a lower mass compact object. 
Although the lifetime of the primary star increases with the decrease of its core mass, 
the fact that the  mass transfer started late in the main sequence phase does not 
significantly affect the lifetime of the star (new $\tau_{\rm ms}=4.0$ Myr). 

The mass transfer continues while the primary is crossing the HG (Case B), transferring mass 
at a high rate $\sim 10^{-2} \mpy$ such that the entire H-rich envelope is removed and
the primary becomes a massive naked helium star with $M_1=10 \msun$.  The secondary, on the 
other hand,  remains on the  main sequence, but with its mass increased to $M_2=60 \msun$. 
Mass transfer ceases as massive helium stars are characterized by small radii ($R_1=0.9 
\rsun$ for the primary).  Thereafter, the primary evolves through the helium star phase, 
losing more mass via stellar winds and finally exploding in a Type Ib supernova. At the 
time of explosion ($t=4.8 Myr$) the primary has a mass of $M_1=6.6 \msun$ with core mass 
of $4.9 \msun$. Such a star forms a massive neutron star ($M_{\rm ns}=1.9 \msun$) as little 
matter is expected to fall back in such a case (e.g., Fryer \& Kalogera 2001). The explosion 
disrupts the system provided that either a high natal kick or a kick off the orbital plane 
is applied, releasing the neutron star from the binary. The massive secondary soon ($t=6.1$ Myr) 
completes its evolution and forms a solitary black hole of $M_{\rm bh}=7.9 \msun$.

{\em Physical properties. }
In Figure~1 we show the initial mass distribution of very young pulsar progenitors. We note 
that the majority of the progenitors are massive $M_{\rm zams}=50-80 \msun$ since high initial 
masses are required in the framework of the above formation channel. Stars must be sufficiently 
massive to promptly complete their evolution within 5 Myr, but not too massive for otherwise 
black hole formation occurs. 

The mass distribution of the very young solitary pulsars formed at $t_{\rm form} \lesssim 5$ 
Myr is also illustrated in Figure~1. As expected these pulsars are very massive: $M_{\rm 
ns} = 1.8-2.5 \msun$. The high mass limit reflects our assumption on the maximum neutron 
star mass ($M_{\rm ns,max}=2.5 \msun$), whereas the peak at $M_{\rm ns}=1.8 \msun$ 
originates from a wide range of progenitor masses. Specifically, our model for single
stars has approximately a bimodal distribution of neutron stars masses;
single stars with $M_{\rm zams}=8-18 \msun$ form $M_{\rm ns}=1.3 \msun$,
while stars with $M_{\rm zams}=18-21 \msun$ form $M_{\rm ns}=1.8 \msun$. 
Such a distribution was adopted following the study of Timmes, Woosley \& 
Weaver (1996), which showed bimodality in the core mass of massive stars reflecting 
the differences in the entropy of their core regions.  \footnotetext{All 
details of the adopted scheme for neutron star mass calculation are given 
in Belczynski et al. (2008).} Therefore, since our scenario preferentially
selects the most massive stars that form neutron stars the peak exists  
at $M_{\rm ns}=1.8 \msun$, while there are virtually no neutron stars with $M_{\rm
ns}=1.3 \msun$.    

In Figure~2 the cumulative distribution of neutron star formation times is 
displayed with solitary neutron stars originating from binaries denoted
by the solid line.  Here, the fraction of neutron stars that have formed before 
a given time is shown. Neutron star formation begins at $t_{\rm form} \sim 4.3$ 
Myr, and is complete by $t_{\rm form} \sim 60$ Myr. The early phase of neutron star formation
from very massive stars proceeded through the mass loss associated with the Roche 
lobe overflow episodes (as described above), while the neutron stars with the 
longest formation times result from the evolution of intermediate mass stars 
either through electron capture supernovae (e.g. Podsiadlowski et al. 2004) or via 
rejuvenation.  For the adopted age of Westerlund 1 $t=5$ Myr we find that the 
fraction of neutron stars formed is $f_{\rm form}=0.001$, implying that for $1000$  
neutrons formed via disrupted binaries, one would form within the first $5$ Myr. Since 
{\em (i)} most of the massive stars in Westerlund 1 appear to reside in binaries and 
{\em (ii)} most of the massive binaries ($\sim 95\%$; Belczynski, Lorimer \& Ridlay 2008) 
are disrupted in supernovae explosions, one requires at least $1000$ massive stars 
($\gtrsim 8 \msun$) in the Westerlund 1 cluster to form one very young pulsar. 
One should note the steepness of the cumulative curve for the times close to the 
age of Westerlund 1, where the fraction changes by more than two orders of magnitude 
from $t_{\rm form}=4.6$ Myr ($f_{\rm form}=0.0001$) to $t_{\rm form}=5.9$ 
($f_{\rm form}=0.01$). 

There are about $\sim 10^5$ stars in Westerlund 1 and only stars with masses 
exceeding $2 \msun$ have had sufficient time to contract onto the main sequence. 
If we adopt a power-law IMF with an exponent of $-2.7$, there are about $10^4$ 
stars more massive than $\gtrsim 8 \msun$ that can potentially form a neutron star. If 
these $10^4$ massive stars are required to form the very young pulsar as observed 
in Westerlund 1 it would yield a fractional efficiency of 0.0001, corresponding 
to $f_{\rm form}$ for a cluster age of 4.6 Myr. It should be noted that this approximate 
estimate is not intended to provide a calibrated rate/efficiency of young pulsar 
formation, but only to illustrate the an order of magnitude consistency of our results 
with the particular observation of one object.

\subsection{Single star progenitor}
 
The possibility that a primordial single star forms a very young pulsar has also been 
explored.  The calculations were performed with the {\tt StarTrack} code for solar 
metallicity and with standard wind mass loss rates as described in Belczynski et al. (2008). 
Within the framework of our assumptions regarding the relation between initial and final 
compact object masses (Belczynski et al. 2008), stars within the mass range of $M_{\rm zams} 
\sim 8-21 \msun$ can form neutron stars if the maximum neutron star mass is taken to be $2.5 
\msun$. The minimum formation time of a neutron star is $t_{\rm form}=9.4$ Myr for the most massive 
progenitor ($\sim 21 \msun$) and is longer by nearly a factor of 2 than the time required 
to produce a pulsar in Westerlund 1. This is shown  in Figure~2 where the cumulative 
distribution of the neutron star formation times for primordial single stars (dashed line) 
is also shown. 

Adopting a higher maximum neutron star mass of $3.0 \msun$ results in a slight decrease of 
the minimum formation time to $t_{\rm form}=9.0$ Myr, corresponding to the final evolution of a $M_{\rm 
zams}=21.5 \msun$ star. This is due to the fact that our predicted initial-final mass relation 
(see Fig.~1 of Belczynski et al. 2008) steeply rises with an initial mass above $M_{\rm zams}=20 
\msun$ as expected for the increased amount of fall back during compact object formation. 

A further decrease in the minimum time for formation of a neutron star could be accomplished by 
increasing the metallicity of the stars in Westerlund 1.  For example, given $M_{\rm ns,max}=3.0 
\msun$ and allowing for a metallicity of $Z=1.5 \times Z_\odot=0.03$ as may be expected for a 
young cluster, the minimum time would decrease further to $t_{\rm form}=7.7$ Myr 
($M_{\rm zams}=24.0 \msun$), but still significantly high for the age of Westerlund 1. 

The only remaining alternative (within the framework of our adopted stellar models) is to 
increase the stellar wind mass loss rates. In this case, the mass loss rates may reduce the 
stellar mass sufficiently that even initially very massive stars form neutron stars rather than 
black holes.  To examine this possibility, we have introduced two scaling factors to multiply our 
standard wind mass loss rates. Specifically, the factors $f_{\rm wind,H}$ and $f_{\rm wind,He}$ 
are used to scale winds of stars with H-rich envelopes and naked helium stars, respectively. In 
all the calculations presented herein, $f_{\rm wind,H}=f_{\rm wind,He}=1$. 
We follow the Hurley et al. (2000) scheme and the choice of wind mass loss rates. 
In particular, the rates from Kudritzki \& Reimers (1978) and Iben \& Renzini (1993) are 
used for giant branch and further evolution; Vassiliadis \& Wood (1993) are
applied to asymptotic giant branch;  Nieuwenhuijzen \& de Jager (1990) are
used for massive stars, but modified for a metallicity dependence ($\propto 
(Z/Z_\odot)^{1/2}$); Hamman \& Koesterke (1998) is employed for Wolf-Rayet-like 
winds; Humphreys \& Davidson (1994) rates are used for luminous blue
variables ($L>6 \times 10^5$ and $R \times \sqrt{L}>10^5$; where $L,\ R$ are 
luminosity and radius expressed in solar units). 

In Figure~3 the initial-final mass relation for single star models with various wind mass loss 
rates for stars with H-rich envelopes: $f_{\rm wind,H}= 1,\ 2,\ 3$ are presented. We have 
chosen models with high metallicity ($Z=0.03$) and fixed wind mass loss rates for naked helium  
stars $f_{\rm wind,He}=1$.  Since the increased winds reduce the final compact object mass we 
note the change in mass of the most massive star that can still form neutron star.  For $M_{\rm 
ns,max}=3.0 \msun$ it is found that only stars with $M_{\rm zams} \lesssim 24,\ 40,\ 62 \msun$ 
can form neutron stars for $f_{\rm wind,H}=1,\ 2,\ 3$, respectively. It is also found that only 
stars more massive than $M_{\rm zams} \gtrsim 44,\ 50,\ 58 \msun$ can form compact objects within 5 
Myr. Both constraints can only be satisfied for the model with very high winds: $f_{\rm wind,H}=3$. 
Therefore if single star is to produce a very young pulsar the stellar winds would need to be 
increased by factor of $\sim 3$ or higher. 

We find that such a drastic increase is unrealistic as it would prohibit formation of massive 
black holes over $\sim 4.5 \msun$ for a high metallicity and over $\sim 6.5 \msun$ for solar 
metallicity.  However, a few black holes in Galactic X-ray transient sources have dynamical 
mass estimates exceeding $10 \msun$ (e.g., Orosz 2003; Casares 2006). In addition, recent 
observations indicate that stellar black holes can form with even higher masses in low 
metallicity environments; $\sim 16 \msun$ black hole in M33 (Orosz et al. 2007) and $\sim
23 \msun$ black hole in IC10 (Prestwich et al. 2007; see also Silverman \& Filipenko 2008 
for an improved orbital estimate). Although it is expected that black holes form with higher 
mass in lower metallicity environments, it was recently demonstrated that a reduction in 
the wind loss rate was necessary for the formation of such massive black holes (Bulik, 
Belczynski \& Prestwich 2008).

We have only investigated the increase of winds for stars with H-rich envelopes, as it has been 
argued that stellar wind mass loss rates for massive helium stars (i.e., Wolf-Rayet stars) are 
systematically overestimated due to the ``clumpiness'' of mass outflows from these stars (e.g., 
Hamann \& Koesterke 1998; Nugis \& Lamers 2000). The reduction due to the clumpiness is already
included in our study, but a further decrease (factor of $\sim 2$) was recently suggested by 
Vanbeveren, Van Bever \& Belkus (2007), thereby, making it more unlikely that enhanced wind 
loss rates from single stars are responsible for the formation of neutron stars in Westerlund 1.

\section{Discussion}

A population synthesis study of the massive star population for Westerlund 1 has been carried 
out to examine the possible formation history its neutron star member, CXO J164710.2-45526.  
The young age of the cluster ($4 \pm 1$ Myr), corresponding to a star of $35 \msun$ at the 
turn off of the main sequence places important constraints on its prior history.  It is 
found that an isolated neutron star can be formed as a result of the disruption of a binary 
system during a supernova phase (within 5 Myr) provided that the initial progenitor system 
is characterized by primary masses in the range of $50-80 \msun$.  On the other hand, only 
stars of mass $\sim 60 \msun$, for single star models can produce a neutron star within 
5 Myr, however, the rate of mass loss via stellar winds must be enhanced to the extent that 
the formation of black holes with masses exceeding $10 \msun$ becomes problematic. 

A major feature in the binary scenario of neutron star formation is the significant 
mass loss taking place in Case A RLOF. In this mass transfer phase,  a much smaller He core 
($M_{\rm He} \sim 10 \msun$) is formed at the end of the main sequence phase as compared to the 
isolated evolution of the same star ($M_{\rm He} \sim 20 \msun$). The removal of the H-rich 
envelope through the ongoing RLOF stage before central He-ignition (Case B) leads to the 
onset of strong Wolf-Rayet type winds, preventing an increase of the He core mass via H shell 
burning. Although we employ 
revised (decreased to account for ``clumpiness'') wind mass loss rates for naked helium 
stars (e.g., Hamann \& Koesterke 1998), the Wolf-Rayet phase leads to a further loss of $\sim 3 
\msun$ from the primary star. This is a secondary, but nevertheless important effect 
leading to the formation of very young neutron stars.  An additional factor enhancing the 
formation of neutron stars from very massive binary components was pointed out by Brown et 
al. (2001; and references within), who argued that if the H-rich envelope is removed (e.g., 
in case B or early case C RLOF) and H shell burning is extinguished before the end of central 
He burning the resulting Fe core mass is reduced in comparison to the case where the H-envelope
is not removed.  Hence, the formation of a low mass compact object (i.e., a neutron star rather 
than a black hole) could be enhanced. This finding was connected to the supply (or lack thereof 
in the case of the neutron star formation) of fresh helium into the burning core that can alter 
the amount of carbon and thus the carbon burning lifetimes in this phase of stellar 
evolution. Since this effect is not modelled in our simulations, our predictions are more 
conservative. In other words,  we could expect even more massive stars (than predicted here) to 
form neutron stars under the favorable conditions.
 
Due to ejection of matter and natal kicks during a supernova, it is expected that the neutron 
star will be displaced from its formation site.  The projected distance of CXO J164710.2-455216 
from the cluster center is about 2.3 pc (e.g., Muno et al. 2006).  The source could have been 
formed anywhere in Westerlund 1 and been moving since the time of its formation ($t_{\rm form}$) 
with velocity ($V_{\rm ns}$) imposed on the neutron star during the supernova explosion. 
We note that motion of the system prior to the supernova explosion because of interactions with 
other stars or binaries in the cluster could also contribute, 
but the former are expected to be significantly larger.  In Figure~4 the space velocities
of the very young pulsars formed in our simulations under the assumption of a null initial velocity 
of the binary prior to explosion are shown. The results are presented for two calculations; a 
reference calculation with the Hobbs et al. (2005) natal kick distribution ($\sigma=265$ km 
s$^{-1}$), and one obtained with low kicks ($\sigma=133$ km s$^{-1}$).  It is found that the 
average velocities are $V_{\rm ns} = 380,\ 210$ km s$^{-1}$ for the reference and low kick 
models, respectively. In addition, due to the smaller probability of progenitor binary disruption 
(and thus formation of a solitary pulsar) the number of young solitary pulsars decreases from 
$N_{\rm ns} \sim 1200$ for the reference model to $\sim 800$ for the low kick model.  Using the 
velocity distribution with the age of each pulsar (Fig.~2), the average distance traveled since 
its birth is calculated as $D_{\rm ns} = 90,\ 50$ pc for the reference and low kick models, 
respectively.  This reveals that the majority of the young pulsars is bound to leave the cluster.  
For a region characterized by 10 pc (comparable to the size of Westerlund 1) an estimate of the 
probability of finding a pulsar within the cluster yields $\sim 10\%,\ 20\%$ of pulsars 
for the reference and low kick models, respectively.  This 
result may indicate that the predicted fractional efficiency of forming very young pulsars is 
lower (by factors of $\sim 5-10$) than presented in Figure~2. The decreased efficiency (even 
by factor of 10), however, still permits one very young pulsar to be found within the cluster age 
of 5 Myr.

\acknowledgements
We would like to thank K.Stepien and N.Langer for useful comments on
this study. 
KB thanks Academia Sinica Institute of Astronomy and Astrophysics in Taipei
for hospitality. We acknowledge partial support through NSF Grant number 
AST-0703960 (RT),  and by the Theoretical Institute for 
Advanced Research in Astrophysics (TIARA) operated under Academia Sinica and 
the National Science Council Excellence Projects program in Taiwan administered 
through grant number NSC 96-2752-M-007-007-PAE.

\clearpage

\begin{figure}
\includegraphics[width=1.1\columnwidth]{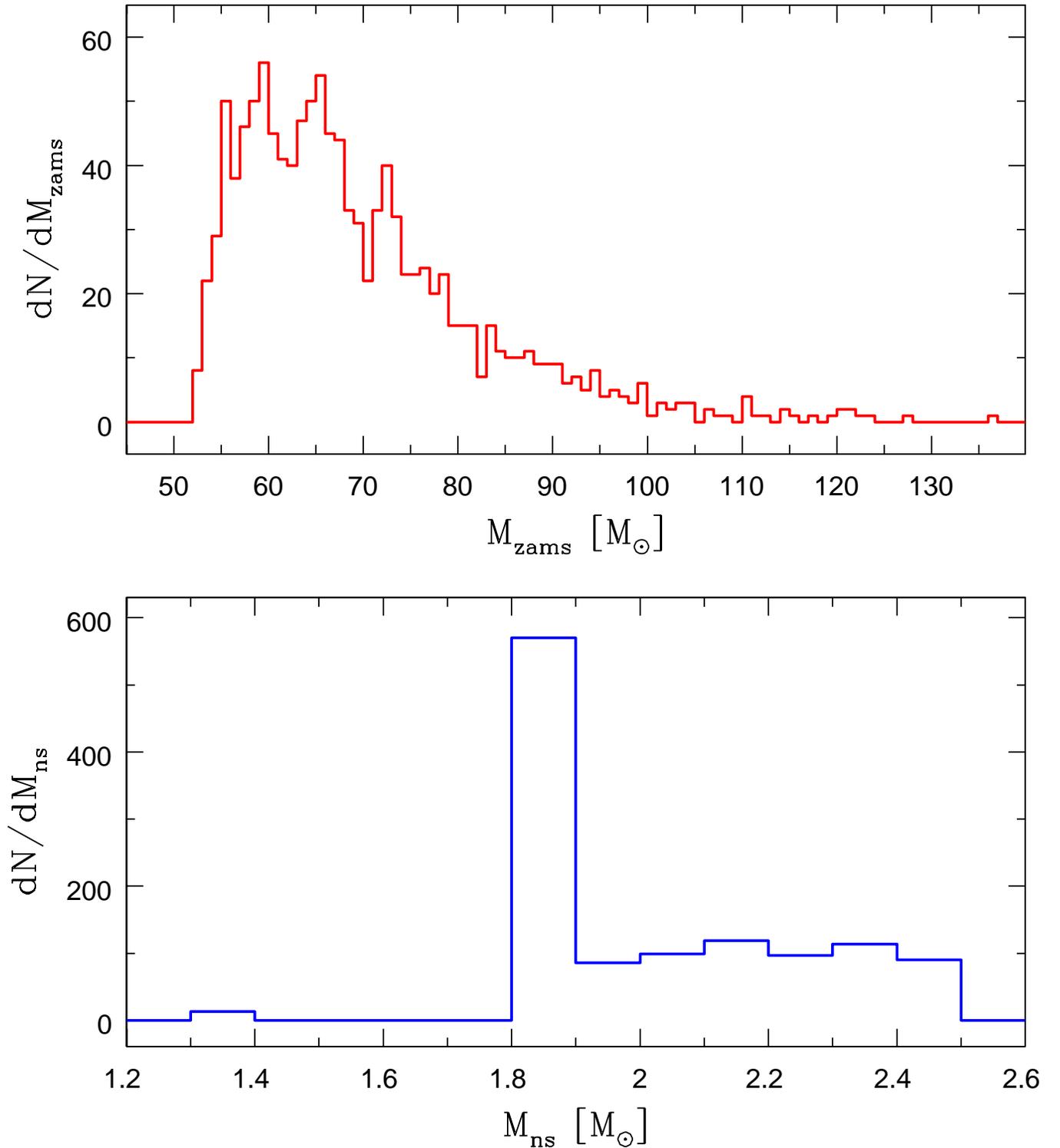}
\caption{
{\em Top panel:} Initial (Zero Age Main Sequence) mass of neutron star progenitors as
predicted in population synthesis calculation for Westerlund 1 at the age of 5 Myr 
(solar metallicity, standard winds). All neutron star progenitors originate from binaries 
that were disrupted in a supernova explosion. Note the very high masses of the progenitors 
($M_{\rm zams} \sim 50-80 \msun$).  {\em Bottom panel:} Neutron star masses predicted in 
the same calculation.  Note the high mass ($M_{\rm ns} \sim 1.8-2.5 \msun$) of the neutron stars. 
}
\end{figure}
\clearpage

\begin{figure}
\includegraphics[width=1.1\columnwidth]{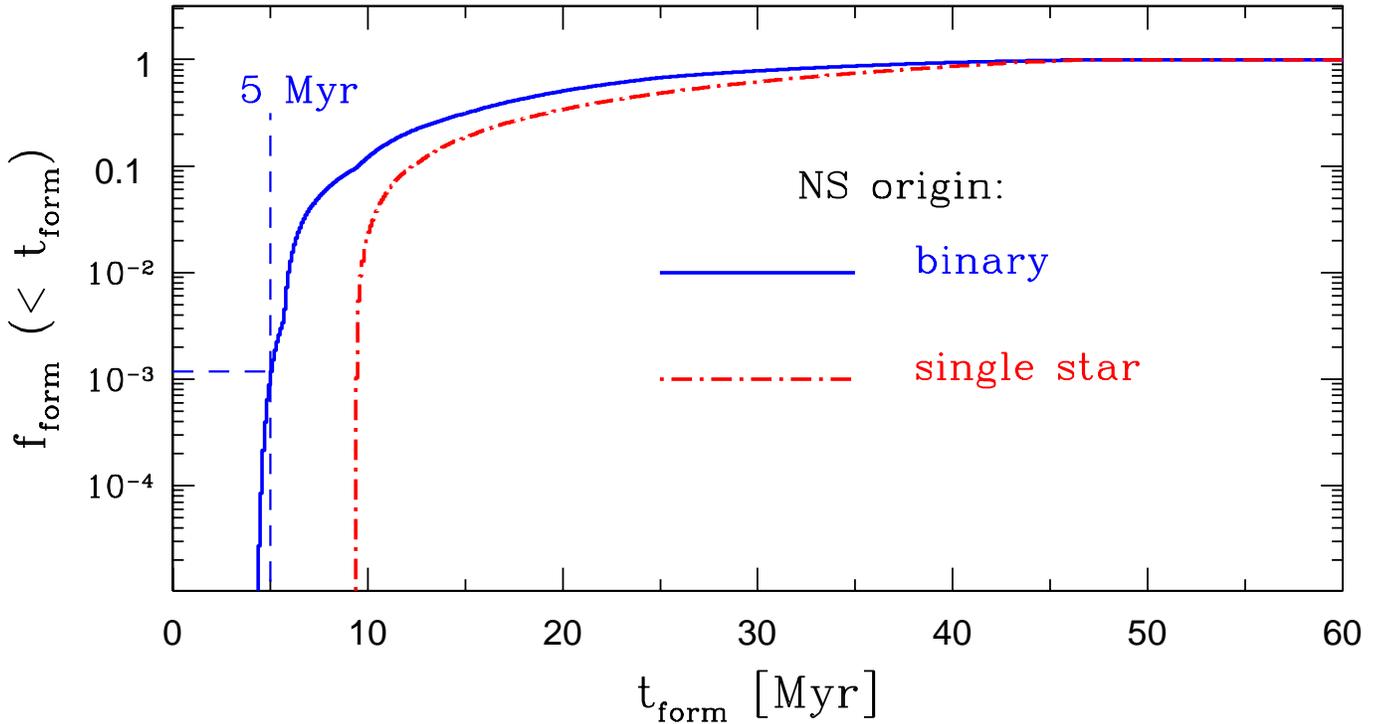}
\caption{
Cumulative distribution of formation time of single neutron stars predicted 
in population synthesis calculation for Westerlund 1. 
Single neutron stars that originate from disrupted binaries are shown with
solid/blue line, while the ones that originate from primordial single stars 
are shown with dashed/red line. 
Note that primordial single stars do not form neutron stars below 9 Myr,
while a small fraction of neutron stars ($f_{\rm form} \sim 0.001$) form in
the first 5 Myr from binary progenitors. 
}
\end{figure}
\clearpage

\begin{figure}
\includegraphics[width=1.1\columnwidth]{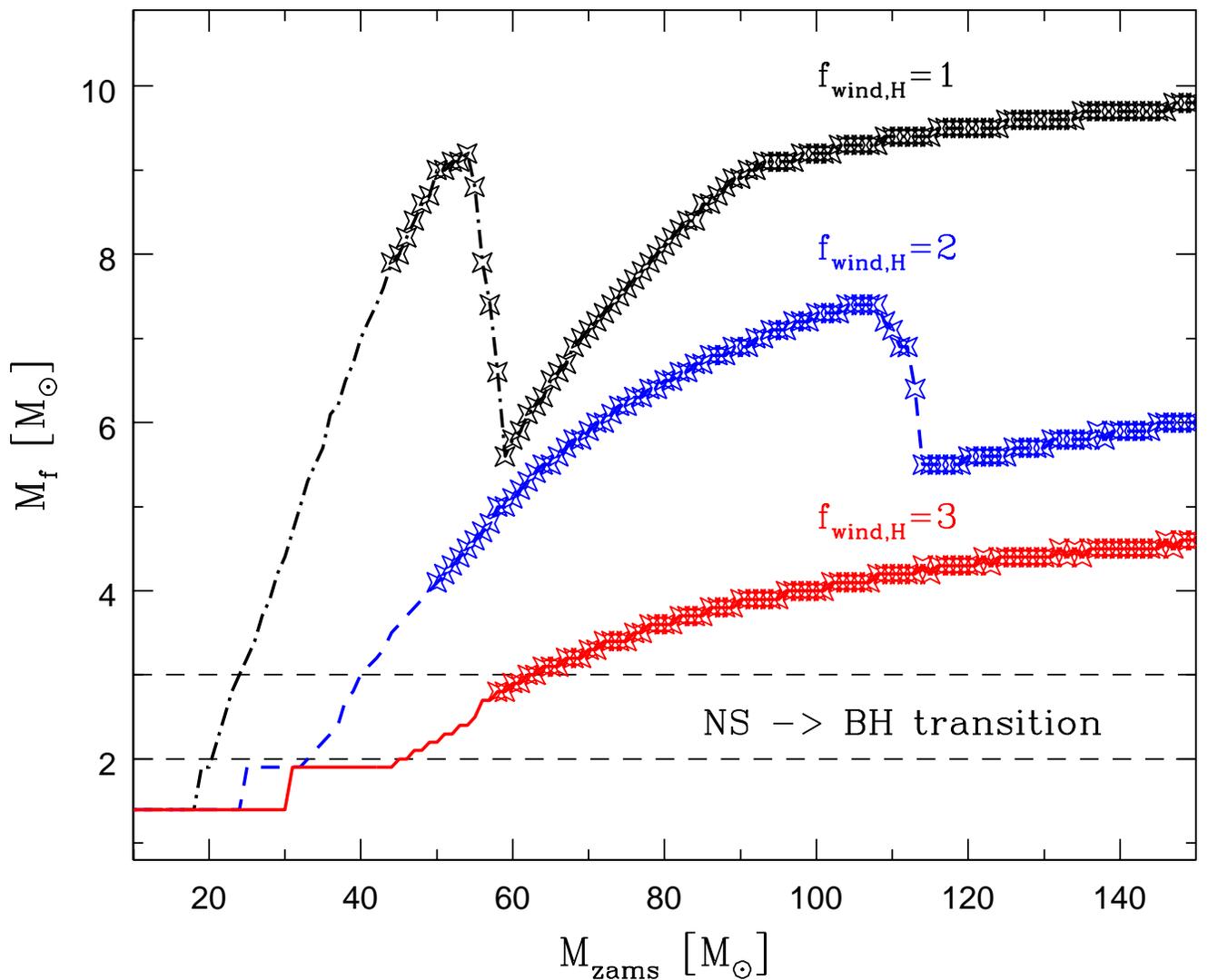}
\caption{
Initial-final mass relation for single star models with various 
wind mass loss rates for stars with H-rich envelopes: $f_{\rm wind,H}=1,\ 2,\ 3$. 
All models are presented for high metallicity ($Z=0.03$) and fixed wind mass loss 
rates for naked helium stars ($f_{\rm wind,He}=1$). We mark the initial mass 
($M_{\rm zams}$) range for each curve (shaded) for which formation of a compact 
object takes shorter than $t_{\rm form}=5$ Myr. Also marked is the final compact 
object mass  ($M_{\rm f}$) range in which it is expected that the transition of a 
neutron star to black hole formation takes place. Note that only for models with 
the very high winds ($f_{\rm wind,H}=3$; bottom curve) can neutron stars be formed 
from single progenitors within $5$ Myr.  
}
\end{figure}
\clearpage

\begin{figure}
\includegraphics[width=1.1\columnwidth]{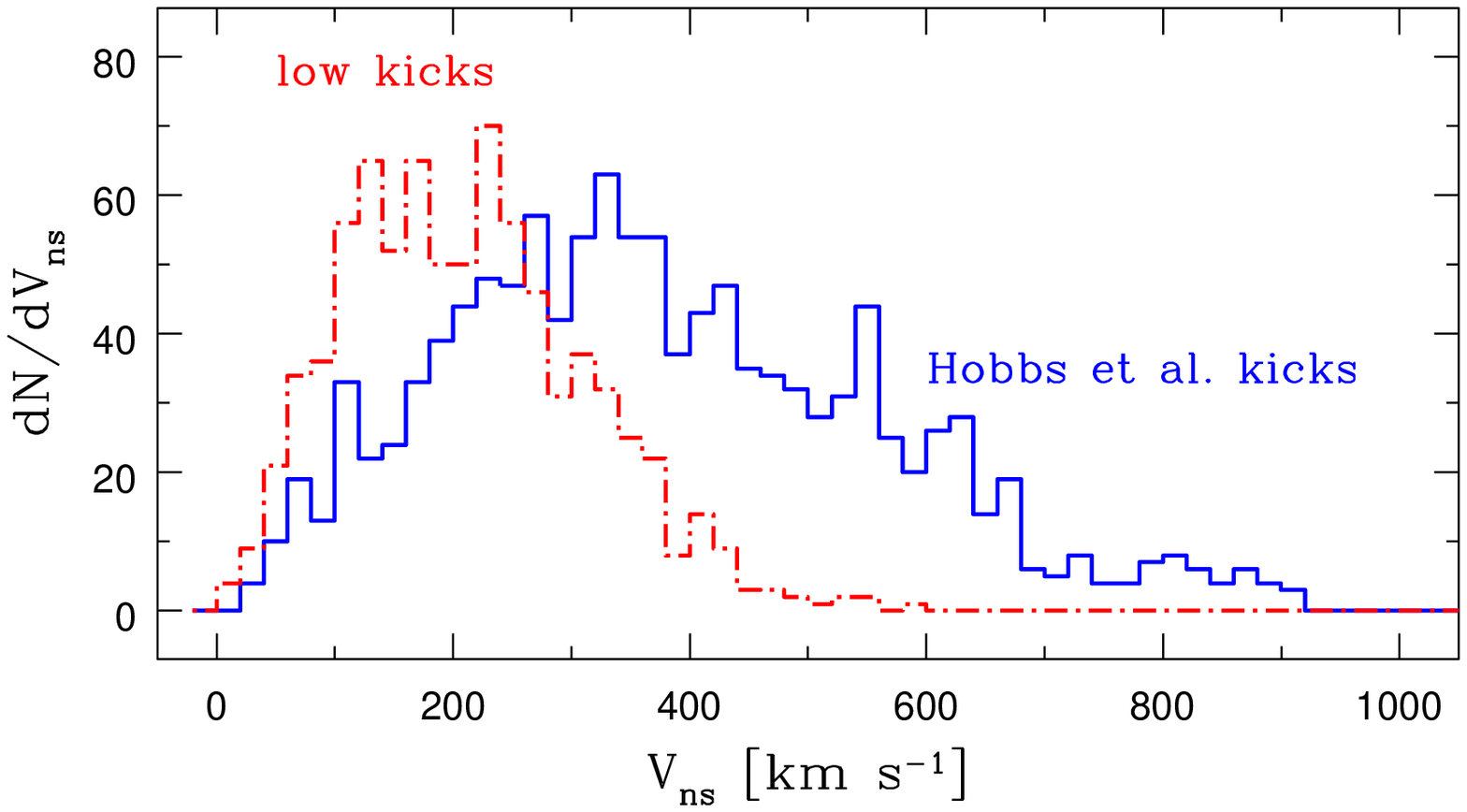}
\caption{
Neutron star space velocities for two natal kick models;
solid line shows results of our reference calculation with Hobbs et al.
(2005) kicks ($\sigma=265$ km s$^{-1}$), while dashed line represents results
obtained with low kicks ($\sigma=133$ km s$^{-1}$).
Note that not only the average velocity decreases from reference ($V_{\rm ns}
\sim 380$ km s$^{-1}$) to low kick model ($V_{\rm ns} \sim 210$ km s$^{-1}$),
but also the number of young solitary pulsars ($N_{\rm ns} \sim 1200$) decreases
due to the smaller probability of progenitor binary disruption for low kicks
($N_{\rm ns} \sim 800$).
}
\end{figure}


\begin{references}

\reference{} Belczynski, K., Kalogera, V., \& Bulik, T.\ 2002a, \apj, 572,
             407 
\reference{} Belczynski, K., Bulik, T., \& Rudak, B.\ 2002b, \apj, 571, 394
\reference{} Belczynski, K., Kalogera, V., Zezas, A., \& Fabbiano, G.\ 2004,
             \apj, 601, L147
\reference{} Belczynski, K., \& Taam, R.\ 2004, \apj, 616, 1159
\reference{} Belczynski, K., Bulik, T., \& Ruiter, A.\ 2005, \apj, 629, 915
\reference{} Belczynski, K., Perna, R., Bulik, T., Kalogera, V., Ivanova,
             N., \& Lamb, D.Q.\ 2006, \apj, 648, 1110
\reference{} Belczynski, K., Taam, R., Kalogera, V., Rasio, F., \& Bulik,
             T.\ 2007, \apj, 662, 504
\reference{} Belczynski, K., Kalogera, V., Rasio, F., Taam, R., Zezas, A., 
             Bulik, T., Maccarone, T., \& Ivanova, N. \ 2008, \apjs, 174, 223
\reference{} Belczynski, K., Lorimer, D., \& Ridlay, J.\ 2008, \apj,
             to be submitted (arXiv:07.......)
\reference{} Brandner, W., Clark, J.,  Stolte, A., Waters, R., Negueruela, I., 
             \& Goodwin, S.\ 2007, \aap, submitted (arXiv:0711.1624)
\reference{} Brown, G., Heger, A., Langer, N., Lee, C., Wellstein, S., \& 
             Bethe, H.\ 2001, New Astronomy, 6, 457
\reference{} Bulik, T., Belczynski, K., \& Prestwich, A.\ 2008, \apj,
             submitted (arXiv:0803:3516)
\reference{} Casares, J. 2006, IAU Symposium 238: "Black Holes: From
             Stars to Galaxies -Across the Range of Masses", in press
             (astro-ph/0612312)
\reference{} Clark, J., Negueruela, I., Crowther, P., \& Goodwin, S.\ 2005, 
             \aap, 434, 949 
\reference{} Clark, J., Muno, M., Negueruela, I., Dougherty, P., Crowther, P., 
             Goodwin, S., \& de Grijs, R.\ 2008, \aap, 477, 147
\reference{} Crowther, P. A., Hadfield, L. J., Negueruela, I., \& Vacca, W. D. 2006, 
             \mnras, 372, 1407
\reference{} Dewi, J., \& Pols, O.\ 2003, \mnras, 344, 629
\reference{} Ergma, E., \& van den Heuvel, E.\ 1998, \aap, 331, L29
\reference{} Fryer, C.\ 1999, \apj, 522, 413 
\reference{} Fryer, C., \& Kalogera, V.\ 2001, \apj, 554, 548
\reference{} Fryer, C., Heger, A., Langer, N., \& Wellstein, S.\ 2002, \apj, 
             578, 335 
\reference{} Hamann W.-R., \& Koesterke L.\ 1998, \aap, 335, 1003
\reference{} Hobbs, G., Lorimer, D., Lyne, A., \& Kramer,
             M.\ 2005, \mnras, 360, 974
\reference{} Humphreys R.M., \& Davidson K.\ 1994, PASP, 106, 1025 
\reference{} Hurley, J.\ R., Pols, O.\ R., \& Tout, C.\ A.\ 2000, \mnras,
             315, 543
\reference{} Iben I.Jr., \& Renzini A.\ 1983, ARA\&A, 21, 271
\reference{} Israel, G. L., Campana, S., Dall'Osso, S., Muno, M. P., Cummings, J., 
             Perna, R., \& Stella, L. 2007, \apj, 664, 448
\reference{} Ivanova, N., Belczynski, K., Kalogera, V., Rasio, F., \&
             Taam, R. E.\ 2003, \apj, 592, 475
\reference{} Kaper, L., Lamers, H., Ruymaekers, E., van den Heuvel, E., \& 
             Zuiderwijk, E.\ 1995, \aap, 300, 446   
\reference{} Kobulnicky, H. A., \& Skillman, E. D.\ 1997, \apj, 489, 636
\reference{} Kudriztki, R., \& Reimers, D.\ 1978, \aap, 70, 22
\reference{} Maeder, A.\ 1992, \aap, 264, 1057 
\reference{} Muno, M., et al.\ 2006, \apj, 636, L41
\reference{} Muno, M. P. et al. 2007, \mnras, 378, L44
\reference{} Nieuwenhuijzen, H., \& de Jager, C.\ 1990, \aap, 231,
             134
\reference{} Nugis, T., \& Lamers, H.\ 2000, \aap, 360, 227
\reference{} Orosz, J.~A\. 2003, in IAU Symposium, Vol. 212, A Massive Star
             Odyssey: From Main Sequence to Supernova, ed. K.~van der Hucht, 
             A.~Herrero, \& C.~Esteban, 365
\reference{} Orosz, J.~A. et al.\ 2007, \nat, 449, 872
\reference{} Podsiadlowski, P., Langer, N., Poelarends, A.J.T., Rappaport,
             S., Heger, A., \& Pfahl, E.D.\ 2004, \apj, 612, 1044
\reference{} Portegies Zwart, S. F., Verbunt, F., \& Ergma, E.\ 1997, \aap, 321, 207
\reference{} Prestwich, A., et al.\ 2007, \apj, 669, L21
\reference{} Silverman, J., \& Filipenko, A.\ 2008, \apj, submitted
             (arXiv:0802.2716)
\reference{} Timmes, F., Woosley, S., \& Weaver, T.\ 1996, \apj, 457, 834 
\reference{} Tout, C., Aarseth, S., Pols, O., \& Eggleton, P.\ 1997, \mnras,
             291, 732
\reference{} Vanbeveren, D., Van Bever, J., \& Belkus, H.\ 2007, \apj, 662, L107
\reference{} Van den Heuvel, E., \& Habets, G.\ 1984, Nature, 309, 598
\reference{} Vassiliadis E., \& Wood P.R.\ 1993, \apj, 413, 641 
\reference{} Webbink, R. F.\ 1984, \apj, 277, 355
\reference{} Wellstein, S., \& Langer, N.\ 1999, \aap, 350, 148 
\end{references}
\end{document}